\documentstyle[aps,preprint]{revtex}

\begin{document}
\begin{center}
{\Large{\bf Liquid-gas phase transition and its order in finite nuclei}}\\

\vskip 1.0cm
Tapas Sil$^1$, S. K. Samaddar$^1$, J. N. De$^2$ and S. Shlomo$^3$\\
$^1$Saha Institute of Nuclear Physics, 1/AF Bidhannagar, Kolkata 700064, India\\
$^2$Variable Energy Cyclotron Centre, 1/AF Bidhannagar, Kolkata 700064, India\\
$^3$The Cyclotron Institute, Texas A$\&$M University, College Station,\\
Texas 77845,USA
\end{center}

\begin{abstract}
The liquid-gas phase transition in finite nuclei is studied in a heated
liquid-drop model where the drop is assumed to be in thermodynamic
equilibrium with the vapour emanated from it. Changing pressure along
the liquid-gas coexistence line of the systems, symmetric or asymmetric,
suggests that the phase transition is a continuous one. This is further
corroborated from the study of the thermal evolution of the entropy
at constant pressure.
\end{abstract}

\vskip 1.0cm
PACS Number(s): 25.70.-z, 21.65.+f, 24.10.Pa
\newpage
The study of liquid-gas phase transition in finite nuclear systems is
of considerable contemporary interest 
\cite{poc,hau,de1,nat1,lee,ell1,gul,das,paw}. Experimental analyses of
the accumulated data on multifragmentation and caloric curves show
compelling evidence of such a transition. Phase transitions are
normally signalled by peaks in the specific heat at constant volume
$C_V$ with rise in temperature. Theoretical models of different
genres, such as the microcanonical \cite{gro} or the canonical 
\cite{bon} description of multifragmentation, the lattice-gas model
\cite{gul,das,sam} or even the microscopic treatment in a relativistic
\cite{sil} or a nonrelativistic \cite{de1} Thomas-Fermi framework
support such a structure in the heat capacity. A clear idea about
the subtle details of the liquid-gas phase transition in finite
nuclei, however, has not emerged yet. Confusion remains about whether
the system has evolved dynamically through the critical point
\cite{ell1,nat2,rad}; a coherent picture about the 
order of the phase transition is also missing. Analyses 
of the EOS group \cite{ell2} and the ISiS group \cite{kle} in the scaling
model give strong circumstantial evidence for a continuous 
(second order) phase
transition. Calculations performed in a mean field model \cite{lee}
also lead to a similar conclusion.
Predictions from the lattice-gas models
\cite{gul,pan} are, however, compatible with a first order transition.

Symmetric infinite nuclear matter is effectively a one-component
system; with heating, it undergoes a first order phase transition.
On the other hand, in case of two-component asymmetric nuclear matter,
as was shown in a comprehensive analysis by M$\ddot u$ller and
Serot \cite{mul} in a relativistic mean-field framework,
the separate conservation of the neutron and
proton number densities leads the system to a continuous phase transition 
over a finite temperature interval.
This is also supported in the nonrelativistic calculations by Kolomietz
{\it et al} \cite{kol}.
Unlike symmetric nuclear matter, even
a finite symmetric nucleus ($N=Z$) behaves like a two-component system
as the Coulomb interaction lifts the isospin degeneracy. A finite
nucleus is then expected to undergo a continuous liquid-gas phase
transition, if at all. The conflicting predictions from the
previous model analyses do not have realistic inputs of nuclear
physics as relevant for a quantum system of interacting fermions
with a short-range nuclear force as also with the long-range Coulomb
interaction. In this communication, we focus on the nature of 
the phase transition once specific features concerning an atomic 
nucleus are properly taken into account.

 For our study, we choose a representative system, namely,
Rhenium with $A$= 186 and $Z$= 75. Such a nucleus is likely to  be 
formed from the reaction $^{124}Sn+^{124}Sn$ at energies 
around 50 MeV per nucleon \cite{xu} after some nucleons have left
out from the reaction zone as preequilibrium particles. We also investigate
$^{150}Re$ to explore the isospin asymmetry effects in phase transition.
The nucleus is viewed as a spherical liquid drop with asymmetry
$X_0$ defined as $(N_0 - Z_0)/A_0$ where $A_0= N_0+Z_0$, is the mass 
number of the total system. At a finite temperature, we assume the 
depleted nucleus to be enveloped by its own vapour and the system
to be in complete thermodynamic equilibrium conserving the total
number of neutrons and protons. The nucleon distributions in the
liquid and gas are assumed to be uniform in each phase. 
This definition allows 
to explore the liquid-gas coexistence region for a finite system
in close analogy with bulk nuclear matter.

In absence of a well-defined way to write the energy density functional of a
finite nucleus in terms of volume, surface, symmetry and Coulomb terms, we
write the free energy of the nuclear system at temperature $T$ in the single
phase as 
\begin{equation}
F=A_0f_{nm}(\rho,X_0)+F_C+F_{surf}
\end{equation}
where $f_{nm}(\rho,X_0)$ is the free energy per particle of infinite
nuclear matter at density $\rho$ with asymmetry $X_0$ at the same
temperature $T$, $F_C$ the Coulomb free energy and $F_{surf}$ the 
temperature and asymmetry dependent surface free
energy. In the liquid-gas coexistence region, the free energy is given by
\begin{equation}
F_{co}=F^l+F^g,
\end{equation}
where the liquid free energy is,
\begin{equation}
F^l= A^l f_{nm}(\rho^l,X^l)+F_C^l+ F_{surf}^l,
\end{equation}
and the free energy of the emanated gas is 
\begin{equation}
F^g= A^g f_{nm}(\rho^g,X^g)+F_C^g.
\end{equation}  
Here $A^l$ and $A^g$ are the number of nucleons in the liquid and the gas
phase, $\rho^l$, $\rho^g$ and $X^l$, $X^g$ are the corresponding density
and asymmetry. The free energy of infinite nuclear matter is evaluated in
the finite temperature Thomas-Fermi framework with a modified
Seyler-Blanchard interaction \cite{de2}. The Coulomb free energies for the
liquid and the gas, $F_C^l$ and $F_C^g$ are calculated corresponding to a
uniform charged sphere and a spherical-shell, respectively. For simplicity,
their mutual interaction is neglected. The surface
free energy of the liquid part $F_{surf}^l$ is taken as \cite{lev}
\begin{equation}
F_{surf}^l=\sigma(X^l,T)(A^l)^{2/3},
\end{equation}
where the surface energy coefficient is 
\begin{equation}
\sigma(X,T)=[\sigma(X=0)-a_sX^2][1+1.5 T/T_c][1-T/T_c]^{3/2}
.
\end{equation}
Here, $\sigma(X=0)=18$ MeV, $a_s=28.66$ MeV and $T_c$, the critical
temperature of the symmetric nuclear matter is 15 MeV. The surface energy
coefficient decreases with density; we neglect its density dependence for
the liquid part. Since the gas density is very low, its surface energy is
neglected. The total surface energy of the liquid is
\begin{equation}
E_{surf}^l=F_{surf}^l+TS_{surf}^l
\end{equation}
where the total surface entropy is given by
\begin{equation}
S_{surf}^l=-\left(\frac{\partial F_{surf}^l}{\partial T}\right)_V
.
\end{equation}
The total entropy $S_0$ of the system is then calculated as
\begin{equation}
S_0=A^ls_{nm}^l(\rho^l,X^l)+A^gs_{nm}^g(\rho^g,X^g)+S_{surf}^l
.
\end{equation}
The per particle entropy $s_{nm}$ of homogeneous nuclear matter is obtained
from the standard mean-field prescription.

The chemical potentials $\mu_n^l$ and $\mu_z^l$ for neutron and proton in
the liquid are given by
\begin{eqnarray}
\mu_n^l&=&\frac{\partial F^l}{\partial N^l},\nonumber\\
\mu_z^l&=&\frac{\partial F^l}{\partial Z^l}.
\end{eqnarray}
The liquid pressure is obtained  from
\begin{equation}
P=\rho^2\frac{\partial (F/A)}{\partial \rho}.
\end{equation}
Similar equations follow for the gas phase.
For thermodynamic equlibrium between the liquid and the gas, the two
chemical potentials and the pressures in both phases must be the same,
i.e, $\mu_n^l=\mu_n^g,\;
\mu_z^l=\mu_z^g$ and $P^l=P^g$.
Along the coexistence line, the mass of the liquid drop changes. For a
chosen $A^l$, the quantities $\rho^l$, $\rho^g$, $X^l$, $X^g$ and $A^g$ are
determined by exploiting the three thermodynamic equlibrium conditions and
the constraints of baryon number and the total isospin conservation:
\begin{eqnarray}
\rho_{n,z}&=&\lambda\rho_{n,z}^l + (1-\lambda)\rho_{n,z}^g\nonumber,\\
\rho X_0&=&\lambda\rho^lX^l+(1-\lambda)\rho^gX^g.
\label{iso}
\end{eqnarray}
Here $\rho$ is the average nucleon density and $\lambda$ is the liquid
volume fraction.

 The isotherms for the nucleus $^{186}Re$ at $T=7,8,9$ and 10 MeV are
displayed in Fig.1. For comparison, the isotherm for nuclear matter with
asymmetry same as that of $^{186}Re$ is also shown at $T=10$ MeV. The
difference between the isotherms for the infinite and the finite system is
not insignificant. Though the Coulomb and surface have opposing effects on
the pressure, the former wins over the latter at this temperature. The
liquid-gas coexistence lines for $^{186}Re$ for the four temperatures
mentioned are shown by the dotted lines. It is seen that the pressure
changes along the coexistence line, as seen earlier in the case of
asymmetric nuclear matter \cite{mul}; the slope of the coexistence lines
also increases with temperature. It is further noted that at a given
temperature and asymmetry, because of the Coulomb effect, the slope of the
coexistence line of a finite nucleus is more compared to that of asymmetric
nuclear matter. The variation of pressure along the coexistence line is a
pointer to a continuous phase transition. Unlike nuclear matter,
the coexistence lines do not extend from the pure gas phase to the pure
liquid phase. At relatively lower temperatures, it is found that 
as the system expands, the size of the liquid drop depletes 
and reaches a minimum
mass beyond which no thermodynamic equilibrium is possible; 
at higher temperatures, 
with compression the liquid drop attains a limiting mass which decreases with
increasing temperature as is evident from the figure. As an example, the
minimum liquid-drop mass at $T=7$ MeV is $A^l=24$ (marked as 
A in the figure); at $T=10$ MeV,
the limiting liquid-drop mass is $A^l=130$ (marked as B). This implies that at
lower temperatures, a gas of a finite number of nucleons when compressed
start nucleating with a minimum mass for the seed in order to remain in
thermodynamic equilibrium. Similarly, at higher temperatures, for the
coexisting finite system, the evaporated gas should contain a minimum
number of nucleons.

The isospin fractionation along the coexistence line for the nucleus
$^{186}Re$ at $T=8 $ MeV is shown in Fig.2. The system has proton fraction
$Y_0=0.403$ (defined as $Y=Z/A$).  
As the system prepared in the gaseous phase is compressed, the
two-phase region is encountered at the point A with the emergence of a
minimum liquid mass ($A^l=20$) at the point B with 
a proton fraction $Y_B$ larger than
$Y_0$. With further compression, 
the gas phase depletes from 
A to C while the liquid phase grows from B to D attaining the total
mass $A_0$ and proton fraction $Y_0$. During compression, the proton
fractions in both phases decrease, but the total proton fraction $Y_0$
remains fixed, as dictated by the conservation of the total isospin 
given by Eq.(\ref{iso}). 
It is evident from the figure that the gas phase is more neutron-rich
compared to the total system while the liquid phase is comparatively
neutron-deficient as also observed experimentally \cite{xu}. 
This feature becomes more prominent with increasing
liquid mass. In order to explore the asymmetry effect on isospin
fractionation in phase transition, the calculated results for the symmetric
system $^{150}Re$ are also displayed in the figure. Contrary to the
relatively neutron-rich nucleus $^{186}Re$ , 
here it is found that the gas phase is
proton-rich. In symmetric nuclei, since $\mu_z$ is greater than $\mu_n$
because of Coulomb interaction, the separation of the gas phase behaves
more like that of proton-rich nuclear matter. The occurrence of liquid-gas
phase transition in symmetric medium-heavy nuclei should then lead to
preponderance of proton-rich isotopes in energetic heavy-ion collisions and
this can be tested in experiments.

The heat capacity per particle $C_V$ for $^{186}Re$ 
at constant volume  (defined as $\left(\frac{d(E^*/A_0)}{dT}\right)_V$, 
where $E^*$ is the total excitation energy of the
system) is displayed in Fig.3 at a representative volume $V=10V_0$ which
can be interpreted as a {\it freeze-out} volume. Here $V_0$ is the normal
volume of the nucleus calculated with the radius parameter $r_0=1.16$ fm. 
A very broad bump in $C_V$ 
with a maximum at $T\sim 10 $ MeV is seen. The system then corresponds
to a liquid part with $A^l$ around 80, the rest of the nucleons
being in the gas phase. This is
contrary to the results in the microscopic mean-field calculations obtained
earlier \cite{de1,sil} at around the same freeze-out volume, where a much
sharper peak was observed when the system
just vaporises completely. At very high temperature, the heat capacity
saturates at 1.5 corresponding to a pure classical gas. 

The thermal evolution of the entropy per particle for the nucleus
$^{186}Re$ at a constant pressure $P=0.06$ MeV fm$^{-3}$ is shown by the
solid line in figure 4. In contrast to a first order phase transition where
the entropy at constant pressure exhibits a discontinuity at a particular
temperature (the phase transition temperature), here the entropy change is
continuous. The noticeable rise in entropy in the temperature range 8.4 to
10.3 MeV is the manifestation of a liquid-gas phase transition in this
temperature domain for the chosen pressure. 
The dashed line in the figure corresponds to the
entropy evolution for the symmetric nucleus $^{150}Re$. This is quite
similar to that of the asymmetric nucleus considered, but markedly
different from that of  symmetric nuclear matter \cite{mul}
which shows a sharp discontinuity at the transition temperature. 
The striking similarity between the two results shows that the finite size
and Coulomb effects are the dominant factors in determining the nature of
phase transition in finite nuclei.
The continuous change of entropy at a constant pressure plead in favour of the
characterisation of the liquid-gas phase transition in atomic nuclei,
symmetric or asymmetric, as a continuous one.

The liquid-gas phase transition in finite nuclei with explicit conservation
of the baryon number and the total isospin has been investigated in this 
communication in a heated liquid drop model.
The peaked structure in the heat capacity,
though broad, signals the occurrence of a liquid-gas phase transition. From
the evolution of entropy at constant pressure, one sees that the transition
occurs over a range of temperatures; this strongly suggests that the
liquid-gas phase transition in a finite nuclear system is continuous.
The simplified assumptions in the model may affect the results somewhat
quantitatively, but the general qualitative features are expected
to remain unaltered. The thermodynamic concepts may not be very
meaningful when the number of particles in one of the phases is very
small, still this model serves as a window to understand the basic
features of liquid-gas phase transition in finite systems.

\newpage
\centerline{\bf Figure Captions}
\begin{itemize}

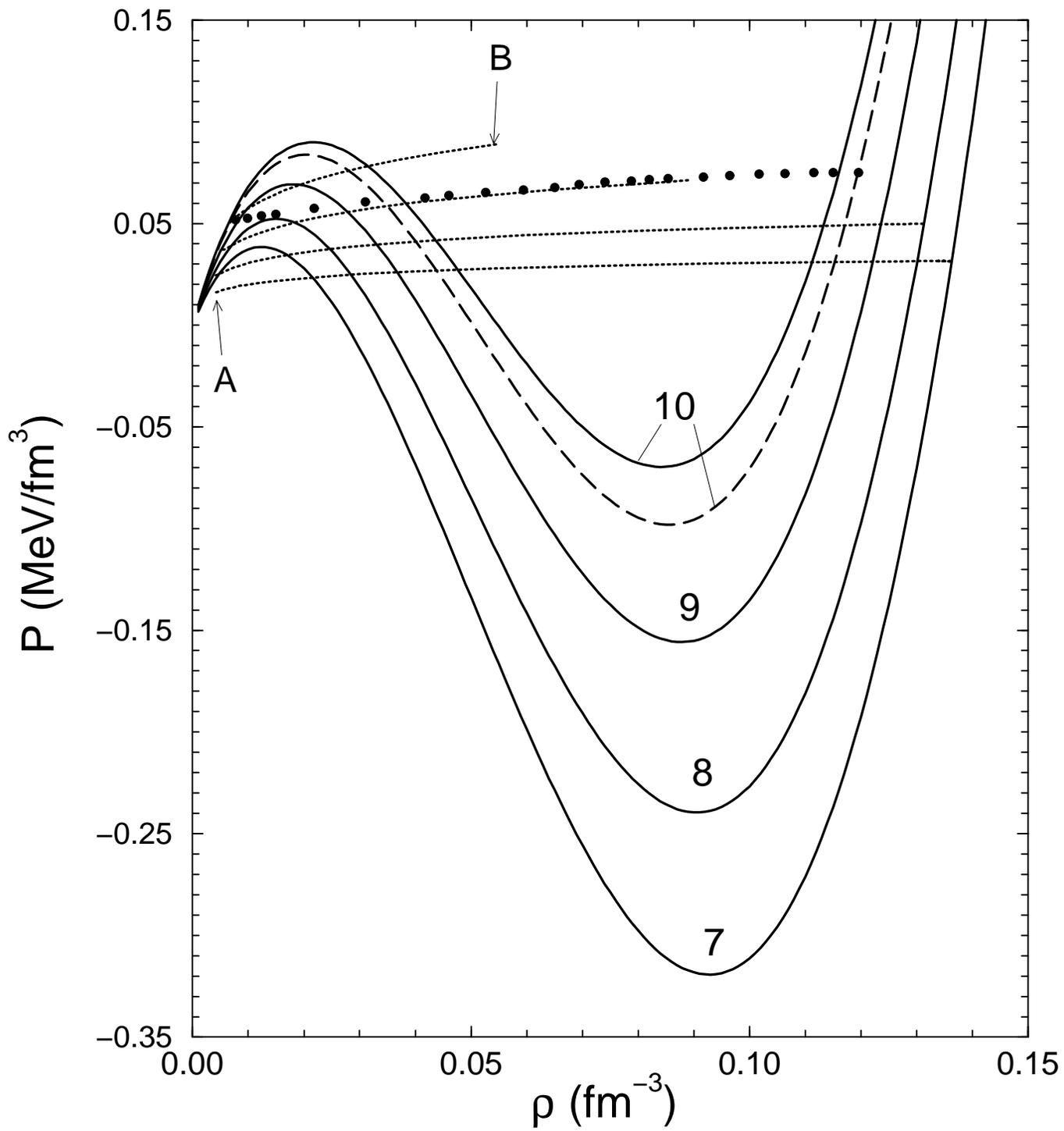
\item[Fig.\ 1] The isotherms for the system $^{186}Re$ at different
temperatures as labelled on each curve are shown as full lines.
The liquid-gas coexistence lines are shown by the dotted lines. 
The long-dashed line refers to the isotherm at $T=$ 10 MeV for 
nuclear matter with asymmetry same as that of $^{186}Re$. The corresponding
coexistence line is shown by the filled circles. 

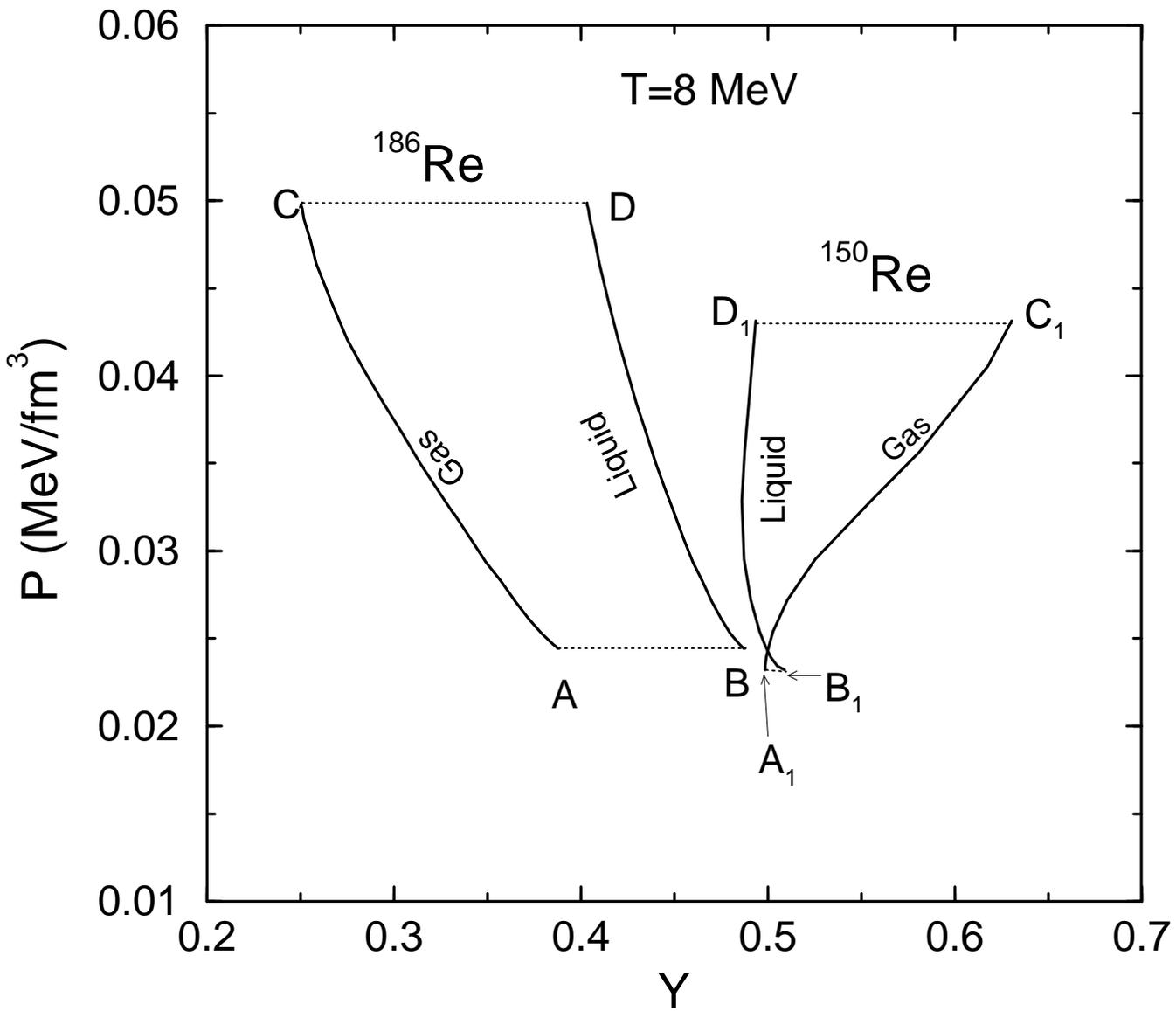
\item[Fig.\ 2] Evolution of proton fraction ($Y$) along the liquid-gas
coexistence line at $T=$ 8 MeV for the systems as shown.

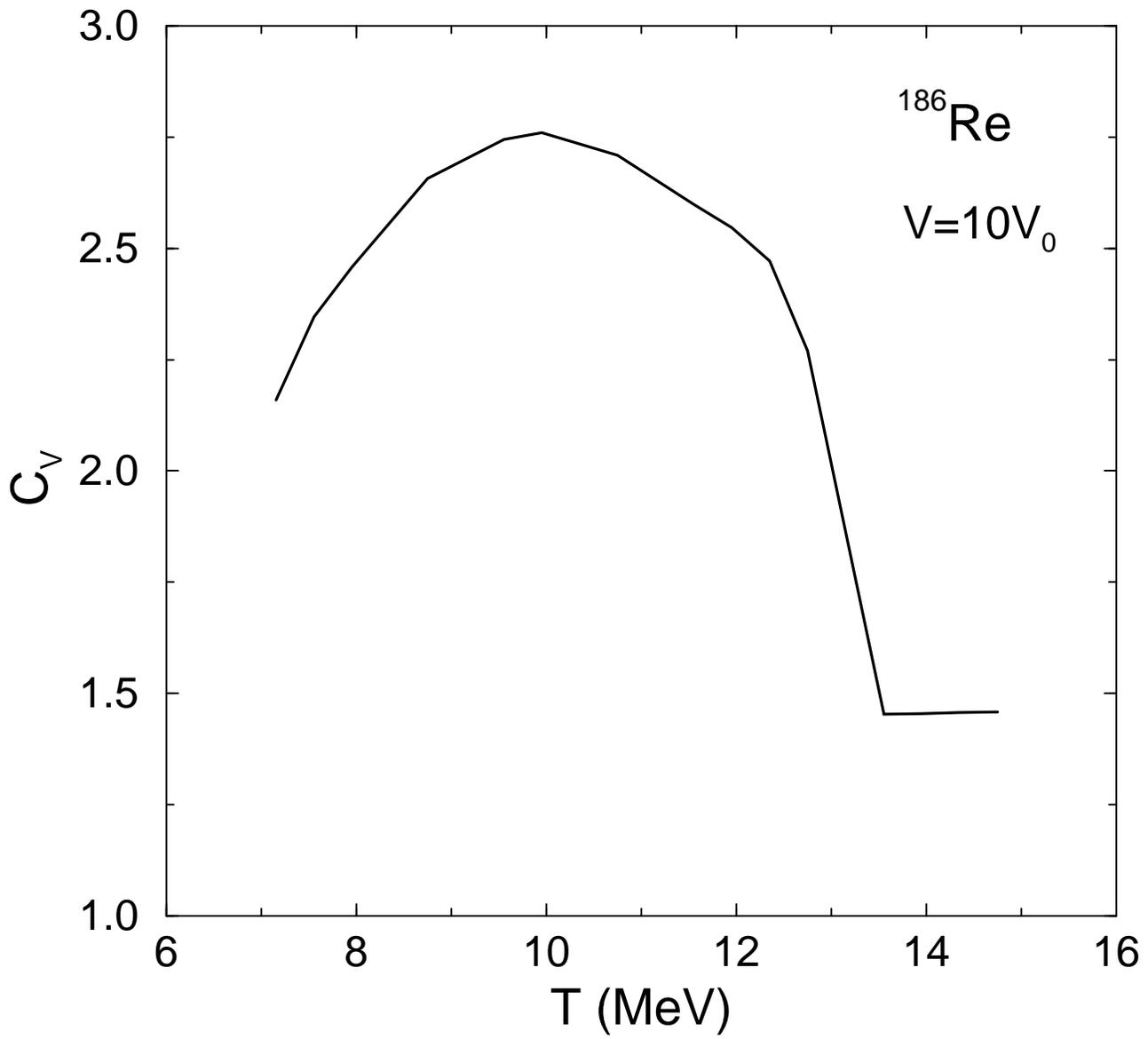
\item[Fig.\ 3] The specific heat capacity $C_V$ as a function of
temperature for $^{186}Re$ in a freeze-out volume 10$V_0$.

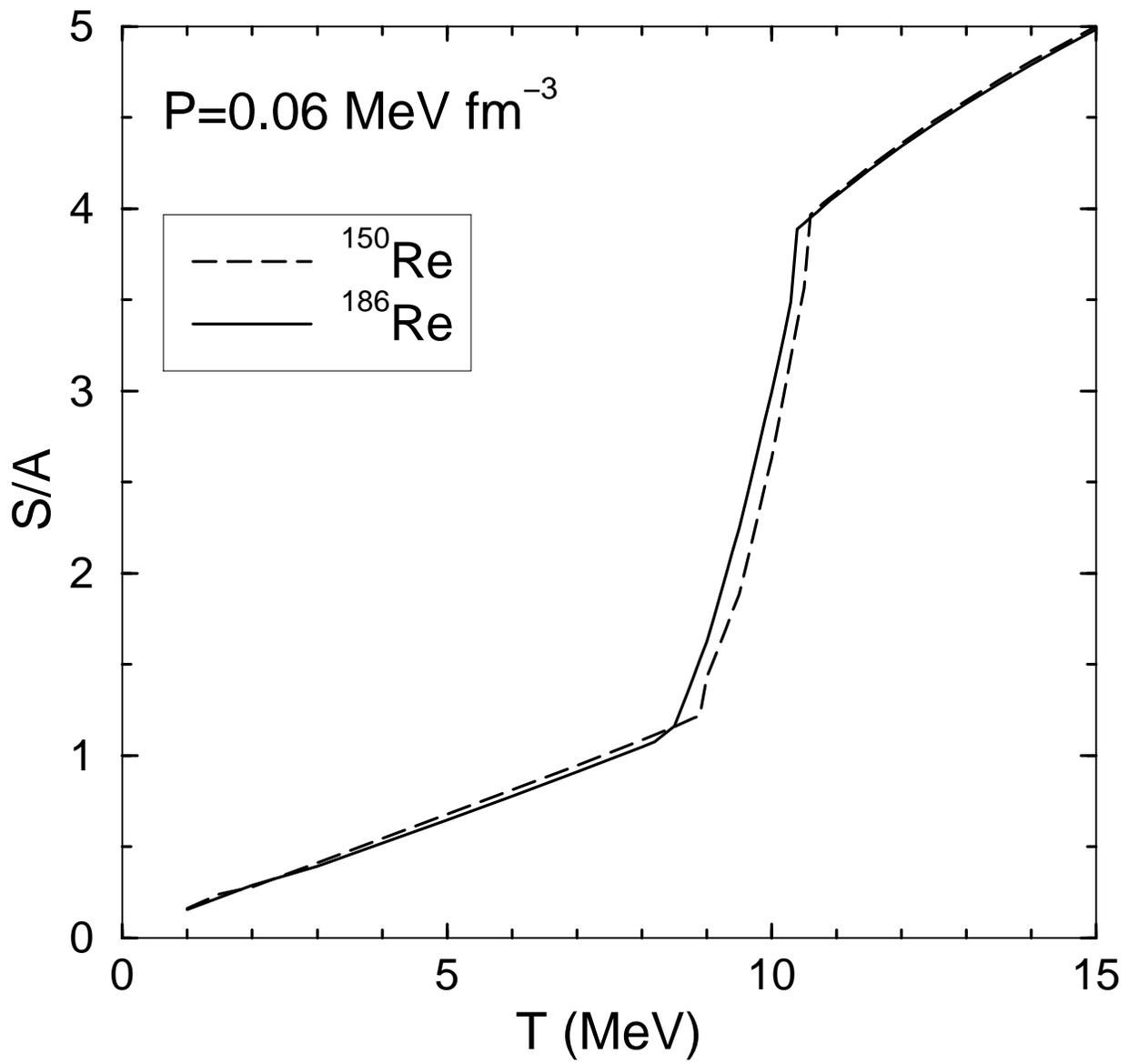
\item[Fig.\ 4] Entropy per particle as a function of temperature at
a constant pressure for the systems $^{186}Re$ and $^{150}Re$.

\end{itemize}
\end{document}